\documentclass[twocolumn,5p]{elsarticle}

\usepackage{lineno,hyperref}
\modulolinenumbers[5]

\usepackage{color}
\usepackage{amsmath}
\usepackage{amssymb}
\usepackage{graphicx}

\begin{document}
\title{Modeling active optical networks}

\author{Giovanni Giacomelli}
\address{Consiglio Nazionale delle Ricerche, Istituto dei Sistemi Complessi, via Madonna del Piano 10, I-50019 Sesto Fiorentino (FI), Italy}

\author{Antonio Politi}
\address{Institute of Pure and Applied Mathematics, Department of Physics (SUPA), Old Aberdeen, Aberdeen AB24 3UE, United Kingdom}

\author{Serhiy Yanchuk}
\address{Institute of Mathematics, Technische Universit\"a{}t Berlin, Strasse des 17. Juni 136, D-10623 Berlin, Germany}
\ead{yanchuk@math.tu-berlin.de}

\begin{abstract}
 The recently introduced complex active optical network (LANER) generalizes the concept of laser system
to a collection of links, building a bridge with random-laser physics and quantum-graphs theory. 
So far, LANERs have been studied with a linear approach. Here, we develop a nonlinear formalism in the 
perspective of describing realistic experimental devices.
The propagation along active links is treated via suitable rate equations, which require the inclusion of
an auxiliary variable: the population inversion.
Altogether, the resulting mathematical model can be viewed as an abstract network, its nodes corresponding
to the (directed) fields along the physical links. The dynamical equations differ from standard network models in that, they are
a mixture of differential delay (for the active links) and algebraic equations (for the passive links).
The stationary states of a generic setup with a single active medium are thoroughly discussed, showing that the
role of the passive components can be combined into a single transfer function that takes into account the corresponding
resonances. 
\end{abstract}


\maketitle

\section{Introduction}

The interest for networks has continuously grown in the last twenty
years. The focus has progressively shifted from the characterization
of their structure towards the study of the underlying dynamics. The
motivation of the massive attention is at least twofold. On the one
hand, many systems of practical and conceptual interest are \textit{de
facto} networks like, e.g., the mammalian brain \citep{Gerstner2014}
and power-grids \citep{Nishikawa2015}. On the other hand several
complex systems can be represented as networks  in suitable abstract
spaces as in the case of climate models \citep{Donges2009a}.

Additionally, networks represent an effective testing ground for theoretical
ideas. Quantum graphs, for instance, offer simplified but non-trivial
models of complex phenonema, such as electron propagation in multiply
connected media, Anderson localization, quantum chaos and even quantum
field theory \citep{Kuchment2008}.

Typically, a network is represented as an ensemble of relatively simple
dynamical systems (the nodes) driven by pairwise interactions schematically
accounted for by a suitable adjacency matrix. This is exemplified
by the Kuramoto model, where the single units are phase oscillators
and the connections are all-to-all \citep{Kuramoto1984}. In some
cases, the connections have their own dynamics, with a massive increase
of the overall computational complexity \citep{gross2008adaptive,Aoki2011,Berner2019}.
This is the case of neural systems, where synaptic plasticity is included
both because of the experimental evidence that the synaptic strength
changes over time and because this mechanism is believed to play a
crucial role in establishing memory \citep{Abbott2000}.

The dynamical properties are so rich that even in identical, globally
coupled oscillators, nontrivial and not yet fully understood regimes
are observed \citep{Zillmer2007}. In this paper, we focus on a class
of networks which naturally arise while considering propagation of
waves of various nature (electromagnetic, acoustic, etc.) through
quasi one-dimensional systems (quantum wires, photonic crystals, thin
waveguides). The novelty and crucial difference with respect to many
other networks is that here the self-sustained dynamical regimes depend
sensitively on the network structure and in particular on the lengths
of the individual  links. Additionally, they offer the possibility
of experimental tests and even the opportunity to develop new devices
such as nanophotonic networks of waveguides \citep{Gaio2019}.

More specifically, this work formalizes the concept of active optical
networks (LANER), going beyond the linear description proposed in
Ref.~\citep{LepriTronoGiacomelli2017,Giacomelli2019}. In our case,
the \textit{network} is a physical network (such as for power grids),
composed of links each characterized by a potentially bi-directional
propagation of electromagnetic waves. \emph{A priori}, there exist
active and passive links (i.e. the electric fields are either damped
or amplified), a little bit like excitatory and inhibitory synaptic
connections in neural systems. A peculiarity of these devices which
distinguishes them from other types of networks is that the wave frequencies
are self-selected and multiple frequencies can coexist. The whole
dynamical structure emerges out of a careful balance and interferences
among the activity along the various links.

This is precisely the reason why it is necessary to include nonlinearities,
as they are ultimately responsible for the saturation of the self-generated
fields, like in standard lasers, a relevant difference being the underlying
complex network LANER structure. In order to keep the model complexity
at a minimum level, without losing physical plausibility, we assume
that the passive links can be all treated as linear damping processes.
As for the active links, the most general approach would require introducing
Maxwell-Bloch equations to account for the spatial structure of polarization
and population along the media \citep{Hess1996}. Given the mathematical
complexity of this type of models, we have restricted our analysis
to active semiconductors-type links, so that the polarization can
be adiabatically eliminated. By following the approach proposed by
Vladimirov and Turaev \citep{Vladimirov2005} for the ring laser,
the spatial dependence of the population dynamics is integrated out
and transformed into a delayed interaction.

As a final simplification, we assume unidirectional propagation along
the active links: this is to avoid the complications arising from
the interactions between counter-propagating waves which would force
us to reintroduce the spatial dependence along the active links.

In Section \ref{sec:LANER-components}, we introduce the mathematical
formalization of the LANER components, starting from the single links
(both active and passive ones) and including the splitters which amount
to a linear coupling between outgoing and incoming fields. The resulting
full LANER network model is introduced in Sec.~\ref{sec:The-LANER-model},
where we show that the most convenient representation consists in
introducing a sort of dual (abstract) network, where the single fields
(with their specific direction of propagation) play the role of nodes,
while the splitters account for the connectivity which is eventually
represented by a nontrivial adjacency matrix. In Sec.~\ref{sec:Networks-single-active}
we consider a general LANER with multiple passive links and a single
active one. The treatment helps understanding that, irrespective of its complexity, the 
passive part can be treated as a single
transfer function, whose resonances contribute to the selection of the relevant frequencies. 

A first exemplification of this approach is the standard ring laser: a single link with a single node.
A less trivial example is discussed in  Sec.~\ref{sec:A-first-non-trivial},
where we discuss a double ring configuration, where only one ring
is  active. In this case, we compute stationary solutions (LANER modes)
and illustrate the effect of the transfer function of the passive
part of the network.  In the last section we summarize the
main results, recall the several open problems and mention possible
directions for future progress.

\section{LANER components\label{sec:LANER-components}}

In this section we outline the mathematical modeling of the main elements of a
LANER: active and passive optical links, and the connecting devices.

\subsection{Link models\label{subsec:Link-models}}


Active links can be realized in several ways, by e.g., laser-pumped erbium-doped fibers or semiconductor amplifiers connected with optical fibers \citep{Franz2008,Leo2010,Mou2012,Herr2014,Bednyakova2015,Romeira2016,Liu2018,Giacomelli2019}.

Importantly for our modeling approach: along the active links we always assume unidirectional 
propagation. Unidirectional propagation avoids dealing with the interaction between counter-propagating 
waves, substantially simplifying the model structure. Additionally, it can be experimentally implemented
by inserting, e.g., optical isolators. 

\begin{figure}
\begin{centering}
\includegraphics[width=0.8\linewidth]{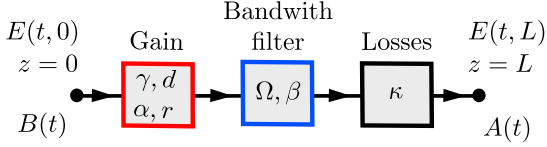}
\par\end{centering}
\caption{\label{fig:link} Schematic representation of the lumped-element approach
for the link containing the gain medium, passive section, and bandwidth
limiting element.}
\end{figure}

We follow the approach introduced in~\citep{Vladimirov2005}, which can be used for
optical systems described by rate equations such as semiconductor
lasers \citep{Soriano2013,Larger2017}. It follows the so-called lumped-element method, where the link
is divided into several components: gain and losses sections, and
the bandwidth limiting element. At variance with~\citep{Vladimirov2005}, here
we do not include the saturable absorber section.
More specifically, let $E(t,z)$ be the slowly-varying amplitude of
the electric field at the position $z$, $E(t,0)$ being
the entrance point and $E(t,L)$ the exit point relative to the propagation
direction -- $L$ is the length of the link (see Fig.~\ref{fig:link}). 
The corresponding propagation time is $T=L/v$, where
the light group-velocity $v$ is assumed to be constant. The direct
application of the approach from \citep{Vladimirov2005} leads to
the following relation between the amplitude of the electric field
at the ends of the link

\begin{equation}
\frac{1}{\beta}\frac{\partial}{\partial t}E(t,L)  =  -\left(1-i\frac{\Omega}{\beta}\right)E(t,L)+\sqrt{\kappa}e^{(1-i\alpha)G(t)/2}E(t-T,0)\label{eq:active link}
\end{equation}
\begin{eqnarray}
\frac{1}{\gamma}\frac{\partial G(t)}{\partial t}  =  d-G(t)+r\left(1-e^{G(t)}\right)|E(t-T,0)|^{2},\label{eq:gain}
\end{eqnarray}
where $G(t)$ is the integral of the local population inversion $n(t,z)$\footnote{At variance with \citep{Vladimirov2005}, for the sake of elegance,
here we time shift the definition of $G$ by $T$.} 
\[
G(t)=\int_{0}^{L}n(t-T,z)dz.
\]

The parameters $\Omega$ and $\beta$ are the central frequency and
the bandwidth of the field filter respectively; $\gamma$ is the carrier
density relaxation rate, $\alpha$ the linewidth enhancement factor;
$d$ is the normalized injection current in the gain section; $\kappa$
accounts for possible additional losses affecting wave propagation;
finally, $r=\frac{vg\Gamma}{\gamma}$, where $g$ is the differential
gain and $\Gamma$ is the transverse modal fill factor.

Since the derivation of the model (\ref{eq:active link},\ref{eq:gain})
follows closely \citep{Vladimirov2005}, it is not reported here.
We limit to explain the role of each term. The differential equation
(\ref{eq:active link}) corresponds to a filter with Lorenzian
line shape $\beta\exp\left[(-\beta+i\Omega)\xi\right]$ of the input signal 
$\sqrt{\kappa}\exp\left[(1-i\alpha)G(t)/2\right]E(t-T,0)$,
where $E(t-T,0)$ is the field entering the link $T$ time before, $\sqrt{\kappa}$
is the total attenuation through  nonresonant linear intensity losses,
and $\exp\left[(1-i\alpha)G(t)/2\right]$ is the amplification and
phase-shift factor due to the semiconductor gain medium. The rate
of change of the gain in equation (\ref{eq:gain}) is proportional
to the injection current $d$, the term $-G(t)$ describes the gain
decay without emission with the rate $\gamma$. The expression 
$r\left(1-\mathrm{e}^{G(t)}\right)|E(t-T,0)|^{2}$
is the contribution of the electric field to the gain rate; it is
proportional to the variation of the intensity during the passage through
the link. For instance, if the electric field is amplified during
the passage, i.e. $|E(t-T,0)|^{2}<|E(t,L)|^{2}$, then this term is
negative, thus contributing to the gain decay.
Notice that the structure of the final model does not depend on the spatial distribution of the pump density, which 
enters the final equation only via the integral $d$.

The minimal, ring configuration, periodic boundary condition $E(t,0)=E(t,L)$,
was already considered in~\citep{Vladimirov2005}.
Here, we discuss true network configurations starting from
the inclusion of passive links, which can be treated as input-output
\begin{equation}
E(t,L)=\mathbf{\sqrt{\kappa}}E\left(t-T,0\right),\label{eq:passive}
\end{equation}
where $T$ is the propagation time. This equation can
be considered as a special case of equation~(\ref{eq:active link})
for $\beta\to\infty$ (infinite bandwidth) and $G=0$. At variance
with active links, passive ones are allowed to be bidirectional, 
since counter-propagating waves do not mutually interfere. 
As we show later, some of the directions of the passive links are not involved in the stationary dynamics, and the corresponding propagating waves decay in time exponentially to zero. However, for the sake of completeness, we prefer to consider the  general case of bidirectional propagations.

\subsection{Connecting the links \label{subsec:Connecting}}

The coupling circuit elements (splitters) transform the incoming electric
fields ($A_{j}(t)$ variables) into suitable output fields $B_{j}(t)$.
Here below, we refer to a specific but commonly used linear four-ports,
$2\times2$ power splitter, described by a scattering $4\times4$
matrix $\mathsf{S}_{\text{sp}}$ with the following properties (see
Fig.~\ref{fig:coupling}):

\begin{figure}
\centering{}\includegraphics[width=0.9\linewidth]{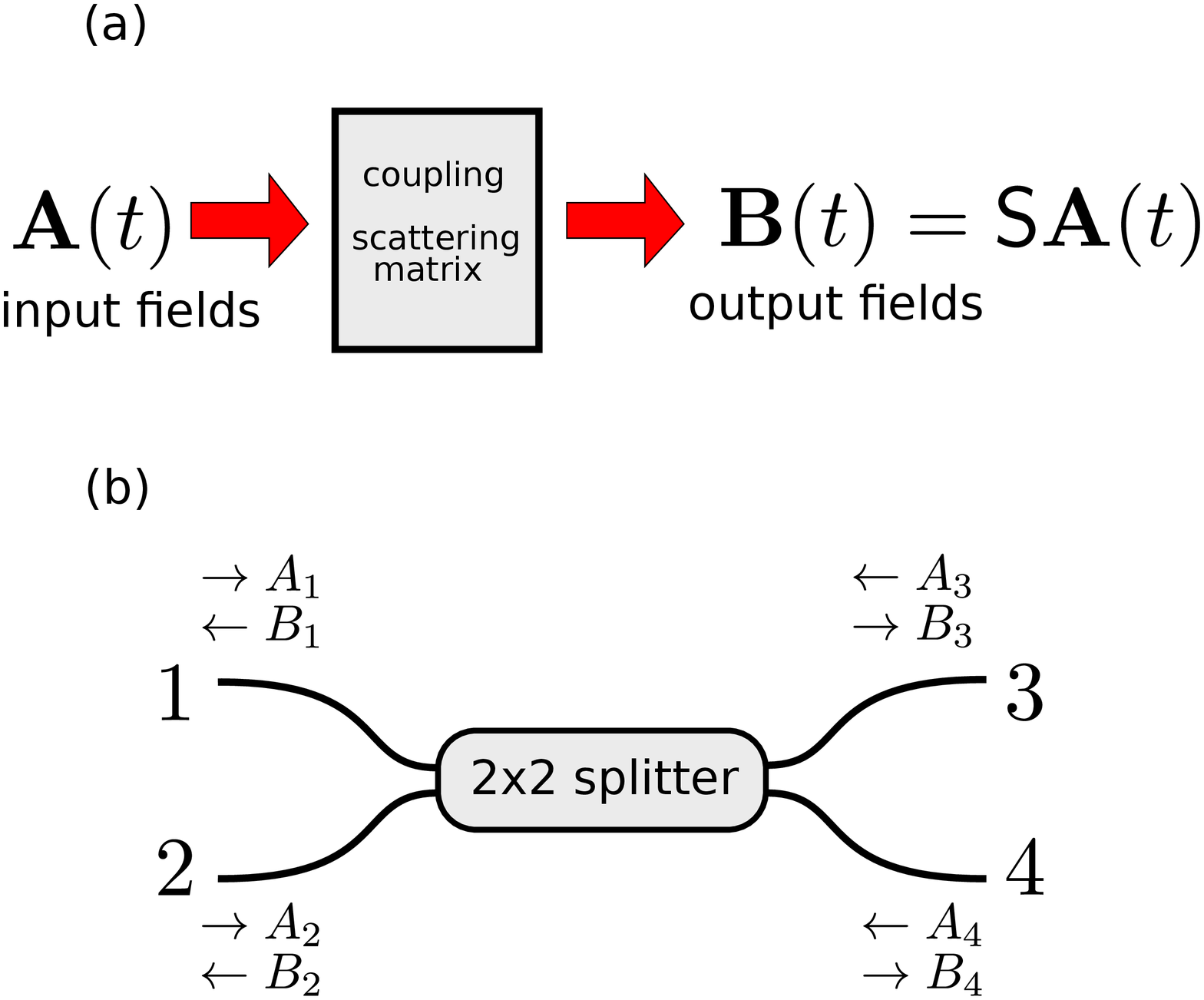} \caption{(a) General linear coupling between the links and (b) a particular
realization with the 2x2 splitter. \label{fig:coupling}}
\end{figure}

\begin{enumerate}
\item Matched, i.e. no reflections, $\mathsf{S}_{\text{sp}}$ has $2\times2$
block-diagonal structure; 
\item Reciprocal (inversion symmetry, $\mathsf{S}_{\text{sp}}$ is symmetric):
$\mathsf{S}_{\text{sp}}^{T}=\mathsf{S}_{\text{sp}}$ (the superscript $T$ denotes the transpose);
\item Lossless (energy conservation, $\mathsf{S}_{\text{sp}}$ is unitary):
$\mathsf{S}_{\text{sp}}^{\dagger}=\mathsf{S}_{\text{sp}}^{-1}$. 
\end{enumerate}
Taking into account the above properties, $\mathsf{S}_{\text{sp}}$
can be written in the form \citep{Pozar2012} 
\begin{equation}
\mathsf{S}_{\text{sp}}=\left(\begin{array}{cc}
0 & \mathsf{s}\\
\mathsf{s}^{T} & 0
\end{array}\right),\label{eq:smatrix}
\end{equation}
where 
\begin{equation}
\mathsf{s}=\mathrm{e}^{i\phi}\left(\begin{array}{cc}
\mathrm{e}^{i\psi_{1}}\cos\theta & \mathrm{e}^{i\psi_{2}}\sin\theta\\
-\mathrm{e}^{-i\psi_{2}}\sin\theta & \mathrm{e}^{-i\psi_{1}}\cos\theta
\end{array}\right).\label{eq:smatrix2}
\end{equation}
Here, $\theta$ measures the splitting ratio, $\psi_{1,2}$ the phase
shifts in the splitting arms, and $\phi$ the overall splitter phase
shifts. The $2\times2$ sub-matrix $\mathsf{s}$ describes the input-output
relation 
\[
\left(\begin{array}{c}
B_{1}(t)\\
B_{2}(t)
\end{array}\right)=\mathsf{s}\left(\begin{array}{c}
A_{3}(t)\\
A_{4}(t)
\end{array}\right)
\]
for the output of the splitter given the input at the right ports
3 and 4, see Fig.~\ref{fig:coupling}. Similarly, $\mathsf{s}^{T}$
transforms the input from the left ports $A_{1}(t)$ and $A_{2}(t)$
into the output $B_{3}(t)$ and $B_{4}(t)$.

\section{The LANER model\label{sec:The-LANER-model}}

\begin{figure}
\includegraphics[width=\linewidth]{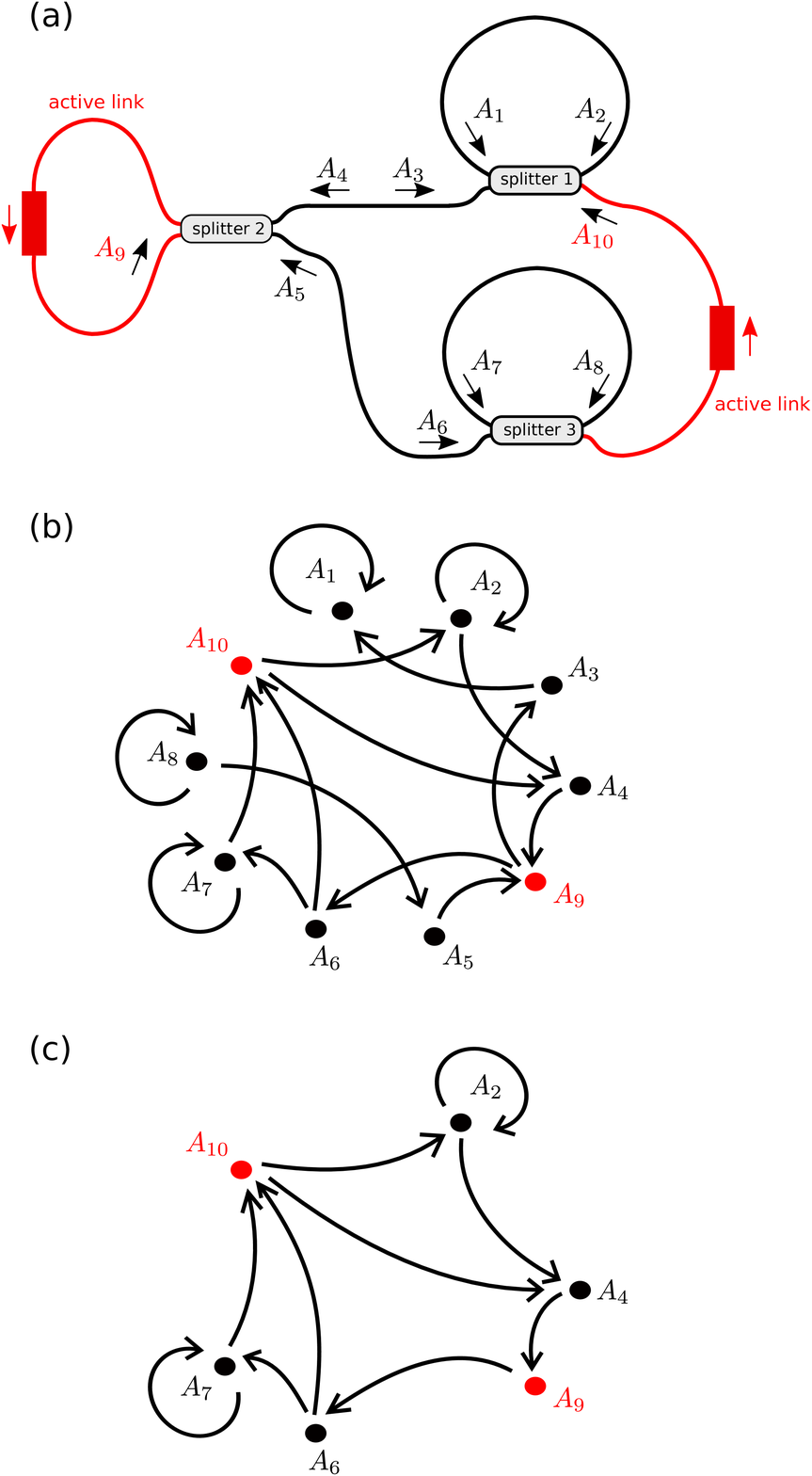} \caption{\label{fig: example} A LANER example. (a) the physical optical network.
(b) the equivalent graph (the abstract LANER network). (c) the reduced
graph determining the dynamics. Active links and fields are denoted in red, while
passive in black.}
\end{figure}

In order to define the LANER model, it is helpful to refer to a specific
example such as the one depicted in Fig.~\ref{fig: example}. It
is also useful to introduce two network representations. The first
one is the physical network $\mathcal{P}$ (see panel (a)), whose
nodes are the splitters, labelled by the index $k$, while the links
are represented by the physical connections between pairs of splitters
(self-connections are allowed), labelled by the index $m$. The second
one, is the ``abstract'' network $\mathcal{A}$, whose
nodes are the fields $A_{j}(t)$ observed at the end of each given
link ($A_{j}(t)=E_{j}(t,L_{j})$, where $L_{j}$ is the length of
the specific link). Hence, each active link is characterized by a
single $A_{j}(t)$ variable (see, e.g. $A_{9}$ and $A_{10}$ in Fig.~\ref{fig: example}(a)).
In passive links, bidirectional propagation is possible; hence we
introduce two variables, $A_{j_{1}}(t)$ and $A_{j_{2}}(t)$ to represent
counter-propagating waves (according to some unspecified rule): see
e.g., the pairs $\left(A_{1},A_{2}\right)$, $\left(A_{3},A_{4}\right)$,
$\left(A_{5},A_{6}\right)$, and $\left(A_{7},A_{8}\right)$ in Fig.~\ref{fig: example}(a).

The links of $\mathcal{A}$ encode the connections among the fields
intervening in each splitter, see Fig.~\ref{fig: example}(b). For
instance, consider node $A_{4}$ of the LANER network. Since the field
$A_{4}$ affects the field $A_{9}$ through the splitter 2, there
is a directed connection from $A_{4}$ to $A_{9}$. Similarly, the
fields $A_{2}$ and $A_{10}$ affect $A_{4}$ through the splitter
1, leading to the connections $A_{2}\to A_{4}$ and $A_{10}\to A_{4}$.
The resulting LANER network for our example is shown in Fig.~\ref{fig: example}(b).

As it will become progressively clear, this latter representation
is more appropriate for the formulation of the dynamical equations.
It is, nevertheless, necessary to introduce a formal relationship
between the two representations. With reference to $\mathcal{A}$,
its nodes can be ordered as we prefer. For the sake of simplicity,
passive nodes (passive links in the physical network $\mathcal{P}$)
are labelled by an index $j\le N_{p}$, while $N_{p}<j\le N$ refers
to the active nodes in $\mathcal{A}$ (active links in $\mathcal{P}$).
Once the ordering has been chosen, the mapping from $\mathcal{A}$
to $\mathcal{P}$ is determined by two functions $M(j)$ and $Q(j)$,
where  $M(j)$ identifies the physical link in $\mathcal{P}$, while
$Q(j)=\pm1$ denotes the corresponding propagation direction (the
value $\pm1$ can be assigned once for all in an arbitrary way but
consistently all over the network). Inversely, given the physical
link $m$ and the propagation direction  $q$, the function $J(m,q)$ determines
the corresponding node within $\mathcal{A}$.

From Eq.~(\ref{eq:passive}), the input-output relationship of  the
passive links are represented as 
\begin{equation}
A_{j}(t)=\sqrt{\kappa_{j}}B_{j}\left(t-T_{j}\right)\;,\quad j=1,\dots,N_{p},\label{eq:passive2}
\end{equation}
where $B_{j}(t)$ denotes the field amplitude at the beginning of
the link (see Fig.~\ref{fig:link}).

The evolution of the active links follows instead from Eq.~(\ref{eq:active link}),
\begin{equation}
\frac{1}{\beta_{j}}\frac{dA_{j}(t)}{\partial t}  = 
-\left(1-\frac{i\Omega_{j}}{\beta_{j}}\right)
A_{j}(t)+\sqrt{\kappa_{j}}\mathrm{e}^{(1-i\alpha_j)G_{j}(t)/2}B_{j}(t-T_{j})
\label{eq:Efinal-3}
\end{equation}
\begin{equation}
\frac{1}{\gamma_{j}}\frac{dG_{j}(t)}{\partial t}  =  d_{j}-G_{j}(t)+r_{j}\left(1-\mathrm{e}^{G_{j}(t)}\right)|B_{j}(t-T_{j})|^{2},
\label{eq:Gfinal-3}
\end{equation}
$j=N_{p}+1,\dots,N$, where the subindex $j$ has been added to the
parameters $\kappa$, $\alpha$, $\beta$, $d$, $\gamma$, $r$, and $T$ to
stress that the various links differ in general from one another.
Two-way propagation in active links cannot be modeled by the above
equations, since the two fields interact with each other. In order
to avoid the additional complications associated with such interaction,
as anticipated,
here in the following we assume unidirectional propagation along 
active links (such as in Fig.~\ref{fig: example}).

\subsection{The scattering matrix}

The equations can be closed by expressing the $B_{j}$ fields 
as functions of the $A_{j}$ variables. This is done, by including
the action of the splitters in the model. Formally speaking, the transformation
is linear and can be written in vector notations as 
\begin{equation}
{\bf B}(t)=\mathsf{S}{\bf A}(t),\label{eq:splitter}
\end{equation}
where ${\bf B}=\left\{ B_{j}\right\} _{j=1}^{N}$, ${\bf A}=\left\{ A_{j}\right\} _{j=1}^{N}$,
while $\mathsf{S}$ represents the scattering matrix, encoding the
action of the splitters.

The structure of $\mathsf{S}$ can be determined from the contribution
of the single splitters. Let $k=1,\dots,K$ be the index numbering
the splitters, and let $a_{1}(k)$ be the input field from the network
$\mathcal{A}$ entering the port 1 of the splitter $k$. Correspondingly,
$a_{2}(k),a_{3}(k),$ and $a_{4}(k)$ are the fields entering the
ports 2, 3, and 4. Similarly, we define $b_{1}(k)$, $b_{2}(k)$,
$b_{3}(k)$, and $b_{4}(k)$ as the outgoing fields from the corresponding
ports of the splitter $k$. The indices of the output fields can be
determined by invoking the mapping between the links of
network $\mathcal{P}$ and the nodes of $\mathcal{A}$,
\begin{equation}
b_{i}(k)=J[M(a_{i}(k)),-Q(a_{i}(k))]~,~i=1,..,4~.
\end{equation}



The relationships resulting in the example in Fig.~\ref{fig: example}
are presented in Table~\ref{tab:splittersexample}. 
\begin{table}
	\centering
\begin{tabular}{|c|c|c|c|c|c|c|c|c|}
\hline 
 & $a_{1}$  & $a_{2}$  & $a_{3}$  & $a_{4}$  & $b_{1}$  & $b_{2}$  & $b_{3}$  & $b_{4}$\tabularnewline
\hline 
\hline 
$k=1$  & 1  & 3  & 2  & 10  & 2  & 4  & 1  & -\tabularnewline
\hline 
$k=2$  & -  & 9  & 4  & 5  & 9  & -  & 3  & 6\tabularnewline
\hline 
$k=3$  & 7  & 6  & 8  & -  & 8  & 5  & 7  & 10\tabularnewline
\hline 
\end{tabular}\caption{\label{tab:splittersexample} Definitions of the input-output fields
for the splitters from the example in Fig.~\ref{fig: example}. $k$
numbers the splitters, $a_{j}$ is the input field into port $j$
and $b_{j}$ the outgoing field from the port $j$. The number is
not defined if the propagation is unidirectional and there is no corresponding
field.}
\end{table}

From the structure of each $\mathsf{S}_{\text{sp}}$ matrix
(see Eq.~(\ref{eq:smatrix})) and the action of all splitters, it follows 
that the elements of the scattering matrix are 
\begin{eqnarray}
S_{j,l} & = & \sum_{k}\Bigl\{[\delta_{j,b_{1}(k)}\delta_{l,a_{3}(k)}+\delta_{j,b_{3}(k)}\delta_{l,a_{1}(k)}]\mathsf{s}_{1,1}(k)+\nonumber \\
 &  & [\delta_{j,b_{1}(k)}\delta_{l,a_{4}(k)}+\delta_{j,b_{4}(k)}\delta_{l,a_{1}(k)}]\mathsf{s}_{1,2}(k)+\nonumber \\
 &  & [\delta_{j,b_{2}(k)}\delta_{l,a_{3}(k)}+\delta_{j,b_{3}(k)}\delta_{l,a_{2}(k)}]\mathsf{s}_{2,1}(k)+\nonumber \\
 &  & [\delta_{j,b_{2}(k)}\delta_{l,a_{4}(k)}+\delta_{j,b_{4}(k)}\delta_{l,a_{2}(k)}]\mathsf{s}_{2,2}(k)\Bigr\}.\label{eq:scattering}
\end{eqnarray}
In simple terms, all elements of the matrix $\mathsf{S}$ are equal
to zero, except those which appear in the $k$th splitter transformation
for some value of $k$. Since each link is assumed to end in a single
well defined splitter, only one of the coefficients of the $\mathsf{s}$
elements can contribute to a given element of the matrix $\mathsf{S}$.

In the case of the LANER depicted in Fig.~\ref{fig: example}, the
matrix $\mathsf{S}$ is 
\[
\mathsf{S}=\frac{1}{\sqrt{2}}\begin{bmatrix}1. & 0. & -1. & 0. & 0. & 0. & 0. & 0. & 0. & 0.\\
0. & 1. & 0. & 0. & 0. & 0. & 0. & 0. & 0. & 1.\\
0. & 0. & 0. & 0. & 0. & 0. & 0. & 0. & -1. & 0.\\
0. & -1. & 0. & 0. & 0. & 0. & 0. & 0. & 0. & 1.\\
0. & 0. & 0. & 0. & 0. & 0. & 0. & -1. & 0. & 0.\\
0. & 0. & 0. & 0. & 0. & 0. & 0. & 0. & 1. & 0.\\
0. & 0. & 0. & 0. & 0. & -1. & 1. & 0. & 0. & 0.\\
0. & 0. & 0. & 0. & 0. & 0. & 0. & 1. & 0. & 0.\\
0. & 0. & 0. & 1. & 1. & 0. & 0. & 0. & 0. & 0.\\
0. & 0. & 0. & 0. & 0. & 1. & 1. & 0. & 0. & 0.
\end{bmatrix}
\]
where we have assumed equal, $50\%$ splitters with zero phase delays,
i.e., $\theta=\pi/4$ and $\phi=\psi_{1}=\psi_{2}=0$.

In the next subsection we derive the dynamical equations, determining
the behavior of the LANER network. However, the network structure
alone provides already important insight in the system properties.
A moment's reflection shows that the network depicted in Fig.~\ref{fig: example}(b)
can be decomposed into three logically concatenated subnetworks, $\mathcal{N}_{1},\mathcal{N}_{2},$
and $\mathcal{N}_{3}$. The subnetwork  $\mathcal{N}_{1}$ is composed of the nodes
$A_{8}$ and $A_{5}$, and it is not influenced by the rest of the network.
Moreover, being entirely passive, its action eventually
vanishes and can be discarded, as shown in Fig.~\ref{fig: example}(c).
Analogously, but with an opposite causality motivation, $\mathcal{N}_{3}$,
composed of the nodes $A_{1}$ and $A_{3}$ does not influence the
remaining six nodes composing $\mathcal{N}_{2}$: it is only forced
by them in a master-slave configuration. 

As a result, the bulk of the evolution arises from the fully
connected component $\mathcal{N}_{2}$~\footnote{The scenario would be slightly more complex in case either $\mathcal{N}_{1}$,
or $\mathcal{N}_{3}$ contain an active link.}.

\subsection{LANER equations}

Eqs.~(\ref{eq:splitter},\ref{eq:scattering}) allow expressing the
field ${\bf B}$ in terms of ${\bf A}$ and, thereby closing the evolution
equations~(\ref{eq:passive2}), which can be rewritten in vector
notations, 
\begin{equation}
\boldsymbol{A_{p}}=\mathsf{K}_{p}\mathsf{P}\mathcal{T}\mathsf{S}\boldsymbol{A},
\label{eq:algeb}
\end{equation}
where $\boldsymbol{A_{p}}=\left[A_{1},\dots,A_{N_{p}}\right]^{T}$
refers to the fields in the passive links, $\mathsf{K}_{p}=\text{diag}\left\{ \sqrt{\kappa_{1}},\dots,\sqrt{\kappa_{N_{p}}}\right\} $
represents the losses in the passive network part, $\mathsf{P}$ is
the projector onto the passive links and, finally, $\mathcal{T}$
is the linear time-delay operator such that $\left[\mathcal{T}\boldsymbol{B}\right]_{j}=B_{j}(t-T_{j})$. 

Similarly, we can eliminate ${\bf B}$ from Eqs.~(\ref{eq:Efinal-3},\ref{eq:Gfinal-3})
and rewrite them as 
\begin{equation}
\mathsf{C}\frac{d\boldsymbol{A}_{a}}{dt}=-\left[\boldsymbol{\text{I}}-i\mathsf{C}\mathsf{\Omega}\right]\boldsymbol{A}_{a}+\mathsf{F}_{1}(\boldsymbol{G})(\mathsf{I}-\mathsf{P})\mathcal{T}\mathsf{S}\boldsymbol{A},\label{eq:diff1}
\end{equation}
\begin{equation}
\mathsf{D}\frac{d\boldsymbol{G}}{dt}=\boldsymbol{D}-\mathsf{F}_{2}(\boldsymbol{G})\left[\left((\mathsf{I}-\mathsf{P})\mathcal{T}\mathsf{S}\boldsymbol{A}\right)\circ\left((\mathsf{I}-\mathsf{P})\mathcal{T}\mathsf{S}\boldsymbol{A}\right)^{*}\right],\label{eq:diff2}
\end{equation}
where $\boldsymbol{A}_{a}=\left[A_{N_{p}+1},\dots,A_{N}\right]^{T}$
are the fields for the active network parts, $\boldsymbol{G}=\left[G_{N_{p}+1},\dots,G_{N}\right]$
are the gains for the active links, $\mathsf{C}=\text{diag}\left\{ \beta_{j}^{-1}\right\} _{j=N_{p}+1}^{N}$
and $\mathsf{D}=\text{diag}\left\{ \gamma_{j}^{-1}\right\} _{j=N_{p}+1}^{N}$
are photon and gain timescales for the active part, $\boldsymbol{D}=\left[d_{j}\right]_{j=N_{p}+1}^{N}$
is the vector of the rescaled injection currents, $\mathsf{\Omega}=\text{diag}\left\{ \Omega_{j}\right\} _{j=N_{p}+1}^{N}$
are frequencies of the spectral filtering for active links, $\mathsf{F}_{1}\left(\boldsymbol{G}\right)=\text{diag}\left\{ \sqrt{\kappa_{j}}e^{(1-i\alpha_j)G_{j}/2}\right\} _{j=N_{p}+1}^{N}$
and $\mathsf{F}_{2}\left(\boldsymbol{G}\right)=\text{diag}\left\{ r_{j}\left(1-e^{G_{j}}\right)\right\} _{j=N_{p}+1}^{N}$
are nonlinear gain functions. Finally, the symbol $\circ$ denotes
the component-wise multiplication.

From the representation (\ref{eq:algeb})--(\ref{eq:diff2}), we
see that the first two equations are linear in $\boldsymbol{A}$ --
the action of the passive part of the network is given by the linear
operator $\mathsf{K}_{p}\mathsf{P}\mathcal{T}\mathsf{S}$ involving
time delays and a matrix multiplication -- while the active subnetwork
is essentially nonlinear. The system also possesses $S^{1}$ symmetry
$\boldsymbol{A}\to\boldsymbol{A}e^{i\varphi}$, $\varphi\in S^{1}$
which is common for laser systems. The resulting system involves delay
differential equations (\ref{eq:diff1})--(\ref{eq:diff2}) as well
as algebraic conditions (\ref{eq:algeb}). Such delay-differential-algebraic
equations appear recently in a model for certain laser systems \citep{Schelte2019}
and they are the subject of emerging theoretical research \citep{Ha2014,Unger2018}.

A peculiarity of the model is the presence of two types of ``oscillators'':
active ones, characterized by two variables, and ``passive'' ones
described by algebraic relations with time-shifts (without time derivatives).
This is reminiscent of Boolean chaos~\citep{GhilZaliapinColuzzi2008,Zhang2009,Rosin2014,LohmannDHuysHaynesEtAl2017,LueckenRosinWorlitzerEtAl2017},
where all equations involve time-shifts and no ``filtering'' by
time-derivatives. Here, only a subset of variables follows such a kind of dynamical evolution. As for the network structure,
the links in $\mathcal{A}$ are all directed: each node can have at
most two outgoing links and two (different) incoming ones.

In the zero delay limit, the evolution equations of the active links (see Eqs.~(\ref{eq:Efinal-3},\ref{eq:Gfinal-3})) reduce to ODEs,
while passive links, determined by the Eq.~(\ref{eq:algeb}), reduce to (linear) algebraic conditions, meaning that they can
be eliminated from the evolution equations. As a result, in this limit, the LANER dynamics is equivalent to a 
network of $N-N_p$ oscillators (the active links) each oscillator being described by three variables (the two components of the
field amplitude, plus the population inversion).
In a sense, the model would be not too dissimilar from ensembles of either Lorenz or R\"ossler oscillators, both characterized by the
same number of variables. An important difference is however in the 
phase-shift symmetry, which is typical for laser models \cite{Uchida2005,Soriano2013}. As a result, the single oscillator cannot be chaotic, and complex dynamics can arise due to interactions. The properties of the zero-delays model are most close to the rate equation models for semiconductor diodes, or compound cavity lasers  \cite{Yanchuk2004,Erzgraber2008,Kominis2017,Erneux2019}.

\section{Networks with a single active link\label{sec:Networks-single-active}}

\begin{figure}
\centering{}\includegraphics[width=0.5\linewidth]{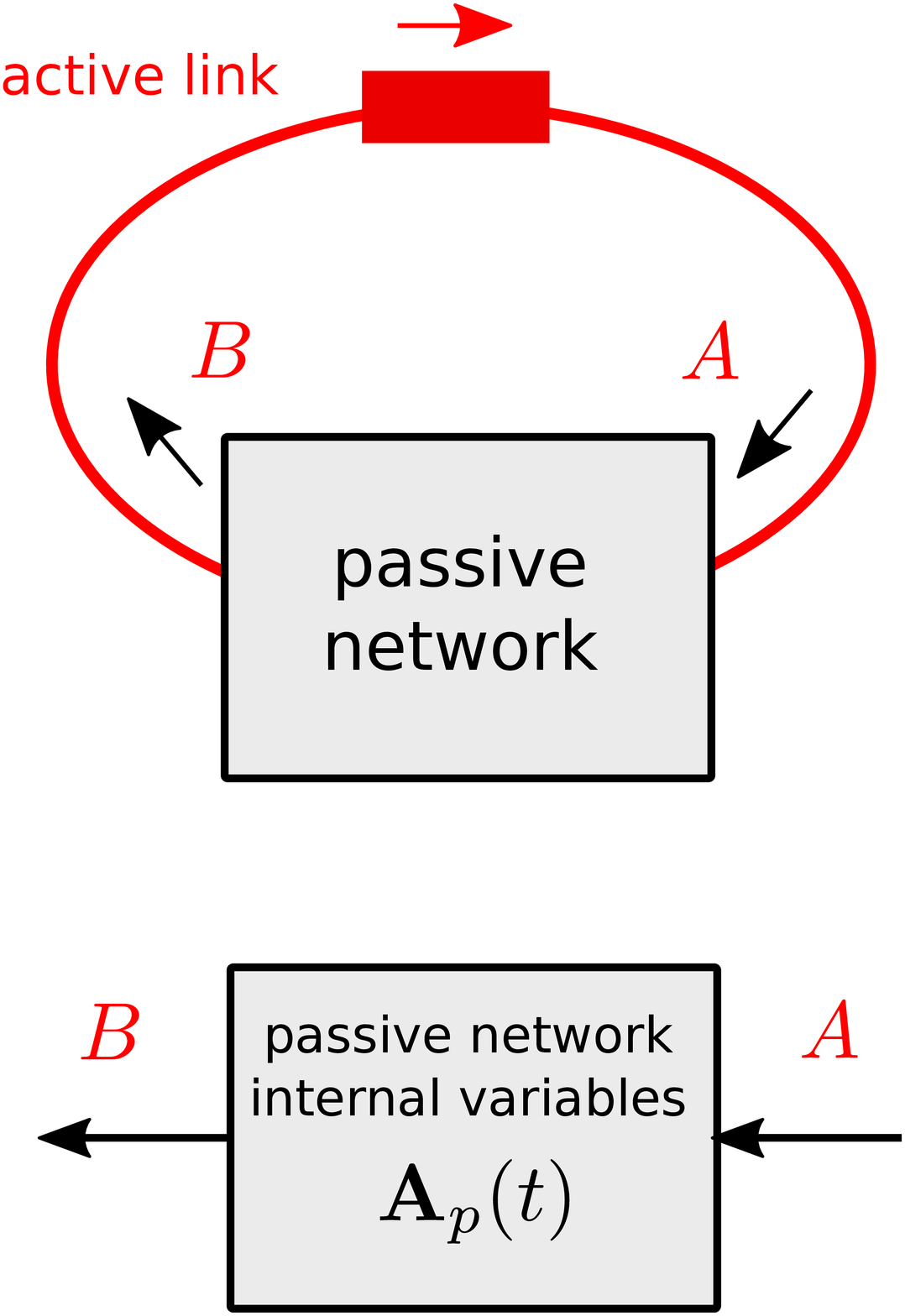}
\caption{Single active medium LANER configuration. \label{fig:Single-active-medium}}
\label{1active} 
\end{figure}

\noindent Because of the presence of nonlinear elements, LANER systems
are expected to exhibit a rich and complex dynamics. In 
order to clarify the role of passive network sections, see
Fig.~\ref{fig:Single-active-medium}, in this section
we analyse setups characterized by a single active (nonlinear)
link. 

Since we have a single active medium, to simplify notations in this
section  we use letters without subscript for the field $A$ (instead
of $A_{N}$), the population inversion $G$, and for all other parameters
of the active link. The scattering (coupling) matrix can be written
 in the block form as 
\begin{equation}
\mathsf{S}=\left[\begin{array}{cc}
\mathsf{S}_{p} & \mathbf{S}_{1}\\
{\bf S}_{2}^{T} & s
\end{array}\right]\label{eq:S-para1}
\end{equation}
where  $\mathsf{S}_{p}$ is a $P\times P$ matrix representing the
connections between the passive links; ${\bf S}_{1}$ and ${\bf S}_{2}$
are two $P$-dimensional vectors encoding the connections between
passive and active links.
Finally the scalar $s$ accounts for the self-interaction present
when the active link is closed onto itself.

As a result, the passive (algebraic) part of the dynamical equation,
Eq.~(\ref{eq:algeb}), can be written as 
\begin{equation}
{\bf A}_{p}=\mathsf{K}_{p}\mathcal{T}\left(\mathsf{S}_{p}\mathbf{A}_{p}+\mathbf{S}_{1}A\right).\label{eq:para1Ap}
\end{equation}

Analogously, the two differential equations describing the active
link are 
\begin{equation}
\frac{1}{\beta}\frac{dA(t)}{dt}  =  -\left(1-\frac{i\Omega}{\beta}\right)A(t)+\sqrt{\kappa}\mathrm{e}^{(1-i\alpha)G(t)/2}B(t-T),\label{eq:para1A}
\end{equation}
\begin{equation}
\frac{1}{\gamma}\frac{dG(t)}{dt}  =  d-G(t)+r\left(1-\mathrm{e}^{G(t)}\right)|B(t-T)|^{2},\label{eq:para1G}
\end{equation}
where  $B(t)$  represents the field amplitude at the beginning of
the active link and is given by 
\begin{equation}
B(t)=sA(t)+{\bf S}_{2}^{T}{\bf A}_{p}(t),\label{eq:para1B}
\end{equation}
see Eqs.~(\ref{eq:splitter}) and (\ref{eq:S-para1}). Expressions
(\ref{eq:para1Ap})--(\ref{eq:para1B}) comprise the complete set
of equations determining the dynamics of the setup in Fig.~\ref{fig:Single-active-medium}.

There is a variety of possible configurations of LANERs with one active
link. Due to the nonlinearities, time-delays, possible complex coupling
structure encoded in the coupling matrix $\mathsf{S}$, the description
of their dynamical properties is interesting but infeasible in complete
generality. Therefore, here we restrict our consideration to the 
solutions with stationary intensity of the form 
\begin{equation}
A=\bar{A}\mathrm{e}^{i\omega t},\quad G=\bar{G},\quad\mathbf{A}_{p}=\bar{{\bf A}}_{p}\mathrm{e}^{i\omega t}\label{eq:stst}
\end{equation}
with time-independent $\bar{A}$, \textbf{$\bar{G}$}, and $\bar{\boldsymbol{A}}_{p}$.
We will call them stationary (LANER) solutions. Substituting (\ref{eq:stst})
into the evolution equations (\ref{eq:para1Ap})--(\ref{eq:para1B}),
we obtain 
\begin{equation}
\begin{array}{cc}
 & \frac{i\omega}{\beta}\bar{A}+\left(1-\frac{i\Omega}{\beta}\right)\bar{A}=\sqrt{\kappa}\mathrm{e}^{(1-i\alpha)\bar{G}/2-i\omega T}\left(s\bar{A}+{\bf S}_{2}^{T}\bar{{\bf A}}_{p}\right),\\
 & \bar{G}=d+r\left(1-\mathrm{e}^{\bar{G}}\right)|s\bar{A}+\mathbf{S}_{2}^{T}\bar{\mathbf{A}}_{p}|^{2},\\
 & \bar{\boldsymbol{A}}_{p}=\mathsf{K}_{\Omega}\left(\mathbf{S}_{1}\bar{A}+\mathsf{S}_{p}\bar{\mathbf{A}}_{p}\right),
\end{array}\label{eq:cweq}
\end{equation}
where $\mathsf{K}_{\Omega}=\text{diag}\left(\sqrt{k_{1}}\mathrm{e}^{-i\omega T_{1}},\dots,\sqrt{k_{n}}\mathrm{e}^{-i\omega T_{n}}\right)$.

The non-lasing state is trivially identified by $\bar{A}=0$,  $\bar{\mathbf{A}}_{p}=0$,
and $\bar{G}=d$. The lasing states can be determined starting from
the last equation, which gives 
\[
\left(\mathsf{I}-\mathsf{K}_{\Omega}\mathsf{S}_{p}\right)\bar{\mathbf{A}}_{p}=\mathsf{K}_{\Omega}\bar{A}\mathbf{S}_{1}.
\]
For physically relevant cases of $\kappa_{j}<1$, $j=1,\dots,N_{p}$,
the matrix $\mathsf{I}-\mathsf{K}_{\Omega}\mathsf{S}_{p}$ is invertible,
hence 
\begin{equation}
\bar{{\bf A}}_{p}=\bar{A}\left(\mathsf{I}-\mathsf{K}_{\Omega}\mathsf{S}_{p}\right)^{-1}\mathsf{K}_{\Omega}\mathbf{S}_{1}.\label{eq:Ap}
\end{equation}
Upon replacing $\bar{{\bf A}}_{p}$ from (\ref{eq:Ap}) in the first
two equations of (\ref{eq:cweq}), we obtain
\begin{align}
 & 1+i\frac{\omega-\Omega}{\beta}=\sqrt{\kappa}\mathrm{e}^{(1-i\alpha)\bar{G}/2-i\omega T}R\left(\omega\right),\label{eq:11}\\
 & \bar{G}=d+r\left(1-\mathrm{e}^{\bar{G}}\right)\left|R(\omega)\right|^{2}\left|\bar{A}\right|^{2},\label{eq:22}
\end{align}
 where the transfer function 
\begin{equation}
R(\omega):=s+\mathbf{S}_{2}^{T}\left(\mathsf{I}-\mathsf{K}_{\Omega}\mathsf{S}_{p}\right)^{-1}\mathsf{K}_{\Omega}\mathbf{S}_{1}\label{eq:R-general}
\end{equation}
accounts for the propagation within the passive part of the network,
including the relative time-delays.

By equating the square moduli of both sides of  Eq.~(\ref{eq:11}), we obtain the expression for the stationary gain
\begin{equation}
\bar{G}(\omega)=\ln\left[1+\left(\frac{\omega-\Omega}{\beta}\right)^{2}\right]-\ln\kappa-2\ln|R(\omega)|.\label{eq:gain-1}
\end{equation}
 Further, Eq.~(\ref{eq:22}) leads to the following stationary field intensity
\begin{equation}
\left|\bar{A}\right|^{2}=\frac{\bar{G}(\omega)-d}{r\left(1-\mathrm{e}^{\bar{G}}\right)\left|R(\omega)\right|^{2}}.\label{eq:A-general}
\end{equation}
Finally, the argument of Eq.~(\ref{eq:11})  yields the condition
for the frequencies 
\begin{equation}
\omega=\frac{1}{T}\left(-\frac{\alpha}{2}\bar{G}(\omega)-\arg\left(1+i\frac{\omega-\Omega}{\beta}\right)+\arg R(\omega)\right)+\frac{2\pi k}{T}, \label{eq:frequency}
\end{equation}
where $k\in\mathbb{Z}$.

The dependence on the delays due to the propagation along the passive links is contained in $\arg R(\omega)$ which is the 
superposition of different ``periods'' $2\pi/T_i$, corresponding to the various passive links. In the limit case of the ring 
laser~\citep{Menegozzi1973} (no passive links), $R(\omega)\equiv 1$.
In the 
next section we consider the less trivial example of a double ring
configuration, illustrating its stationary states.

\section{A first non-trivial LANER: the double ring configuration\label{sec:A-first-non-trivial}}

\begin{figure}
\centering\includegraphics[width=0.5\linewidth]{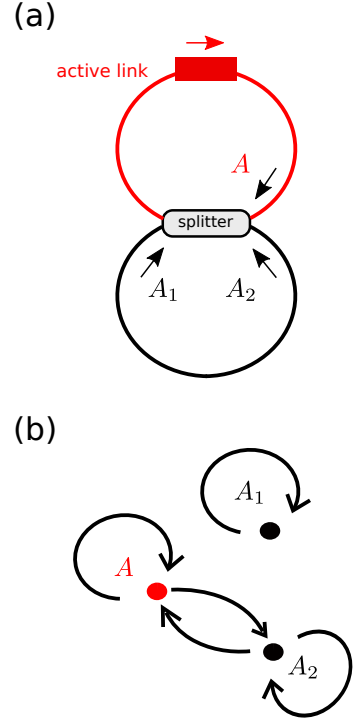} \caption{\label{fig:1para-setup} Double ring LANER with one active and one
passive physical link. Physical network is shown in (a) and the corresponding
abstract network in (b).}
\end{figure}

The simplest, non-trivial example of LANER is the configuration depicted
in Fig.~\ref{fig:1para-setup}, with a single, directed link with propagation delay $T$, accompanied
by a bidirectional passive link with delay $T_2$. 

Similarly to the LANER with one active link in Sec.~\ref{sec:Networks-single-active},
$A(t)$ represents the field at the end of the active link and
$G$ the corresponding gain variable. The equation for the active
link is  
\begin{equation}
\frac{1}{\beta}\frac{dA(t)}{dt}  =  -(1-i\frac{\Omega}{\beta})A(t)+\sqrt{\kappa}\mathrm{e}^{(1-i\alpha)G(t)/2}B(t-T),\label{eq:Efinal-1}
\end{equation}
\begin{equation}
\frac{1}{\gamma}\frac{dG(t)}{dt}  =  d-G(t)+r\left(1-\mathrm{e}^{G(t)}\right)|B(t-T)|^{2},\label{eq:Gfinal-1}
\end{equation}
while for the passive link 
\begin{equation}
A_{2}=\sqrt{\kappa_{2}}B_{2}\left(t-T_{2}\right).\label{eq:passivex}
\end{equation}
The field $A_{1}$ is not coupled with the rest of the network. Since 
  $A_{1}(t)=\sqrt{\kappa_{2}}B_{1}\left(t-T_{2}\right)=\sqrt{\kappa_{2}/2}A_{1}\left(t-T_{2}\right)$,
its amplitude vanishes ($A_{1}(t)\to 0$) exponentially with time and it does not contribute to the stationary regime. Accordingly, such field could be removed from the very beginning in the model.

The coupling is described by the equation
\[
\left(\begin{array}{c}
B\\
B_{2}
\end{array}\right)=\frac{1}{\sqrt{2}}\left(\begin{array}{cc}
1 & 1\\
-1 & 1
\end{array}\right)\left(\begin{array}{c}
A\\
A_{2}
\end{array}\right),
\]
where we assume the splitter action to be 50\% without phase delays.
The final dynamical equations are 
\begin{multline}
\frac{1}{\beta}\frac{dA(t)}{dt}  =  -(1-i\frac{\Omega}{\beta})A(t)+ \\ \sqrt{\frac{\kappa_{3}}{2}}\mathrm{e}^{(1-i\alpha)G(t)/2}\left(A(t-T)+A_{2}(t-T)\right),\label{eq:Efinal-1-2}
\end{multline}
\begin{equation}
\frac{1}{\gamma}\frac{dG(t)}{dt}  =  d-G(t)+\frac{r}{2}\left(1-\mathrm{e}^{G(t)}\right)\left|A(t-T)+A_{2}(t-T)\right|^{2},\label{eq:Gfinal-1-2}
\end{equation}
accompanied by the equation for the passive link 
\begin{equation}
A_{2}(t)=\sqrt{\frac{\kappa_{2}}{2}}\left(-A(t-T_{2})+A_{2}(t-T_{2})\right).\label{eq:passive-1}
\end{equation}
In practice, this equation is a compact way to account for infinitely many delays.

Following the general approach of Sec.~\ref{sec:Networks-single-active},
for the stationary solutions 
\begin{equation}
A_{j}=\bar{A}_{j}\mathrm{e}^{i\omega t},\qquad G=\bar{G}\label{eq:CW}
\end{equation}
the transfer function $R(\omega)$ of the passive part is
\[
R(\omega)=\frac{1}{\sqrt{2}}\left(1-\sqrt{\frac{\kappa_{2}}{2}}e^{-i\omega T_{2}}\left(1-\sqrt{\frac{\kappa_{2}}{2}}e^{-i\omega T_{2}}\right)^{-1}\right).
\]

Given $R(\omega)$, the stationary states can be determined from
Eq.~(\ref{eq:frequency}) for the frequency $\omega$, Eq.~(\ref{eq:gain-1})
for the gain $\bar{G}$, and Eq.~(\ref{eq:A-general}) for the intensity
$|\bar{A}|^2$.

\begin{figure*}
\centering{}\includegraphics[width=1\textwidth]{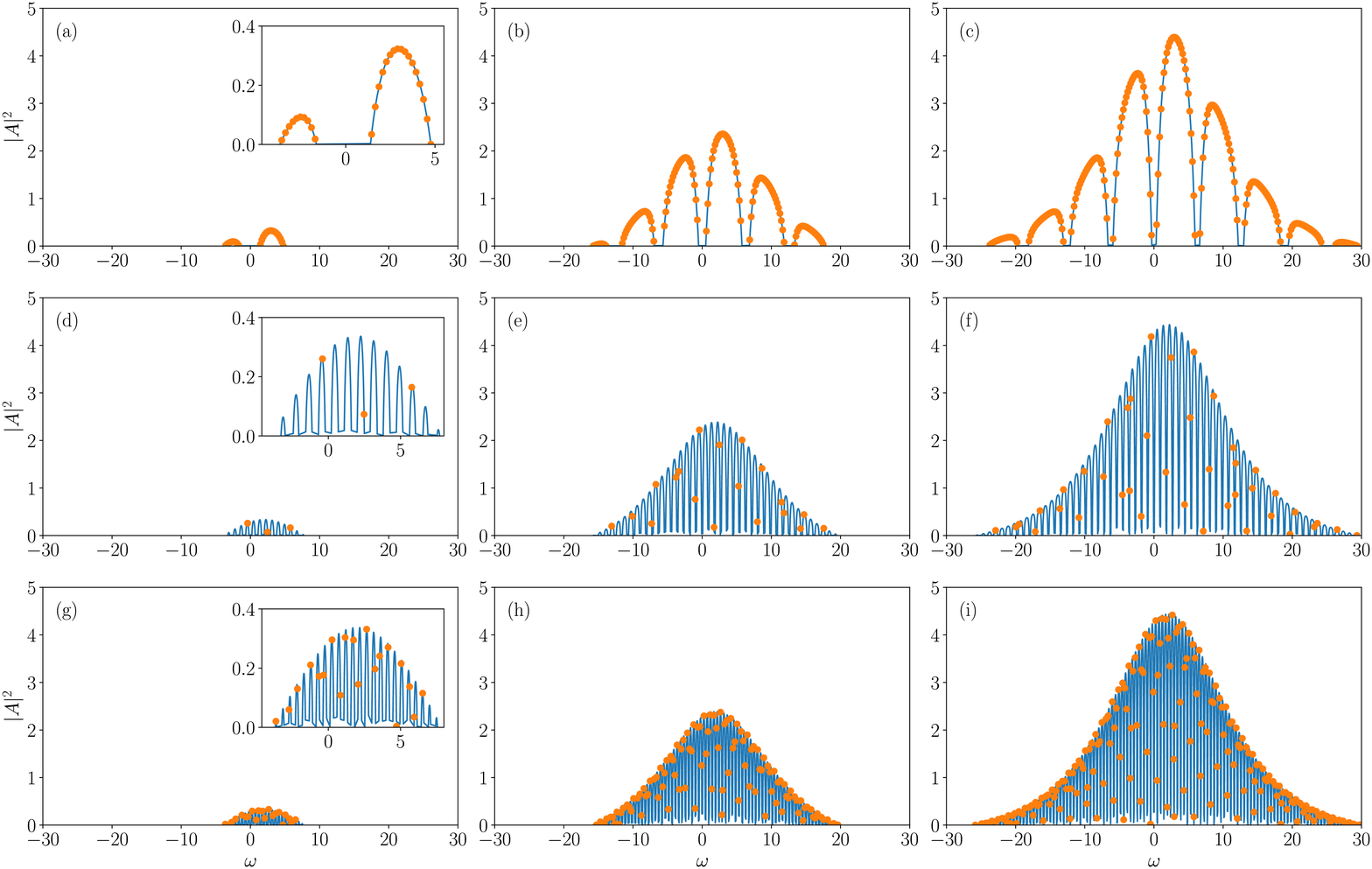} \caption{Stationary states (\ref{eq:CW}) of the double-loop network for different
parameter values. The points refer to the values of the intensity
$|A|^{2}$ of the active link versus its frequency $\omega$. The
blue line shows the dependence of $|A(\omega)|^{2}$ given by Eq.~(\ref{eq:A-general}).
Upper panel (a-c): $T=30$, $T_{2}=1$, and $d=0.5$ (a), $d=1.0$
(b), and $d=1.5$ (c). Middle panel (d-f): $T=2$, $T_{2}=7$, and
$d=0.5$ (d), $d=1.0$ (e), and $d=1.5$ (f). Bottom panel (g-i):
$T=21$, $T_{2}=13$, and $d=0.5$ (g), $d=1.0$ (h), and $d=1.5$
(i). \label{fig:Stationary-states-para1}}
\end{figure*}

A typical set of stationary states is presented in Fig.~\ref{fig:Stationary-states-para1}
for different parameter values. The blue lines show the dependence
of the intensity $|A|^{2}$ in the active link on the frequency $\omega$, as
from Eq.~(\ref{eq:A-general}), irrespective whether the frequency
is supported by the LANER setup. The dots denote the actual stationary
states: the corresponding frequencies are determined by numerically
finding the roots of the scalar  equation (\ref{eq:frequency}).

The various panels are arranged in the following way: rows identify
different sets of propagation (delay) times $T$ and $T_{2}$; columns
identify different values of the pump parameter $d$. More specifically,
the upper row corresponds to a relatively long propagation time along
the active link ($T=30\gg T_{2}=1$); in the middle row, the ratio
is approximately opposite ($T=2<T_{2}=7$); finally, the lower
row corresponds to comparable propagation times ($T=21$, $T_{2}=13$
-- their ratio is an approximation of the golden mean). As for the
columns, from left to right, $d=0.5$, 1, and 1.5, respectively. A
first obvious consideration is that upon increasing the pump amplitude
(i.e. moving from left to right), the number of active modes increases
(see the range of possible frequencies) as well as their amplitude. 
 As in the standard multimode laser, the higher amplitude field modes 
	are located around the optical center frequency $\Omega$. The spectrum is not symmetric around $\Omega$ if the linewidth enhancement factor $\alpha \neq 0$.
By comparing the different rows, we see that the number of
active modes grows with the time-delays. 
Moreover, we notice that the passive section determines the
amplitude of the active modes, inducing a relatively high
sensitivity of their frequency when $T_{2}$ is comparable to $T$.

\begin{figure}
\includegraphics[width=\linewidth]{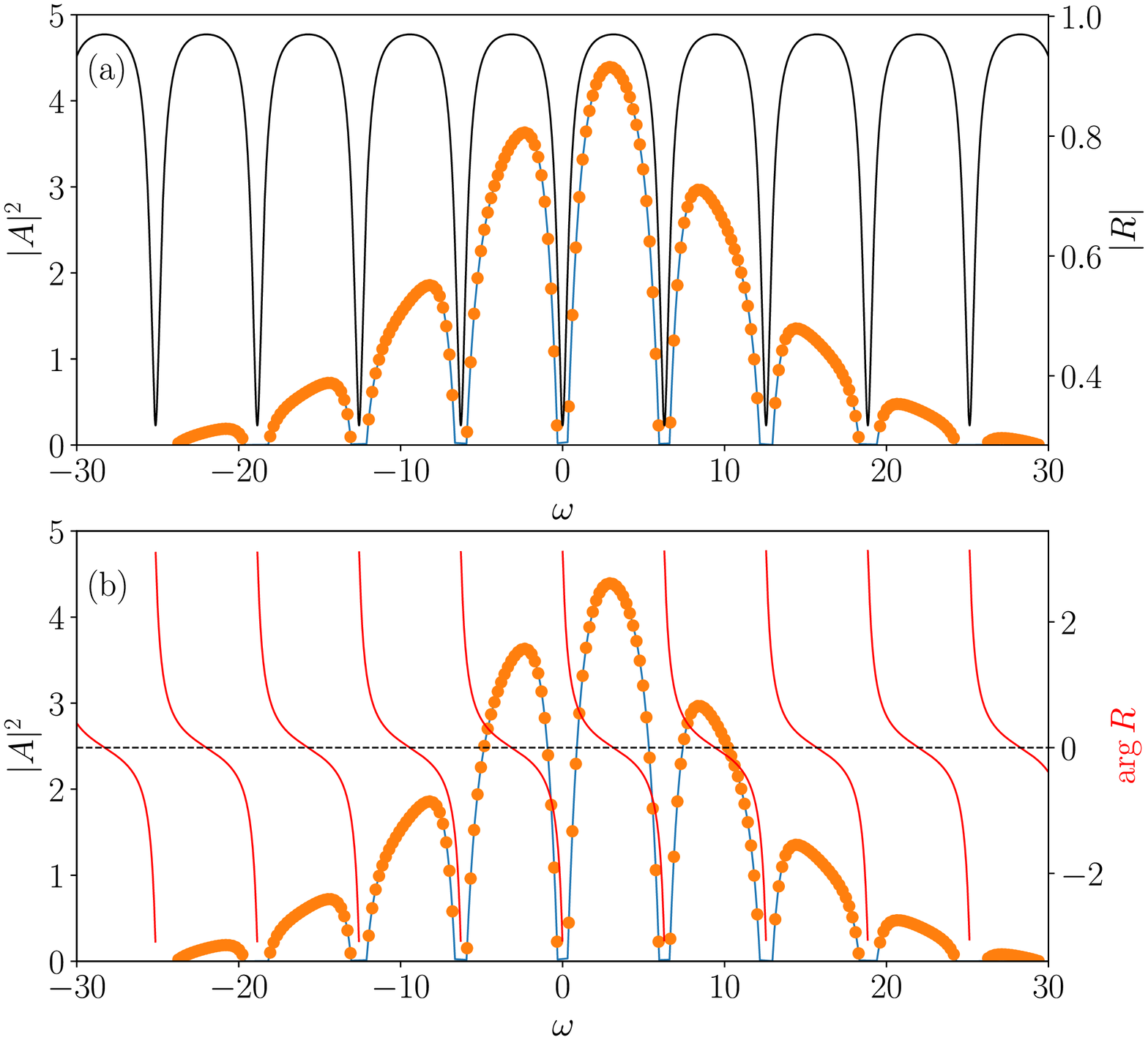}
\caption{\label{fig:R} Modulus $|R(\omega)|$ (panel (a), black line, right
axis) and phase $\arg R(\omega)$ (panel (b), red line, right axis)
of the transfer function of the passive section compared to the intensity
of the field $A$. Parameter values: $T=30$, $T_{2}=1$, and $d=1.5$. }
\end{figure}

The overall scenario can be traced back to the structure of the transfer function $R(\omega)$.
In Fig.~\ref{fig:R}, we separately plot modulus (upper panel) and phase (bottom panel) of $R(\omega)$
for the same delays as in the first row of Fig.~\ref{fig:Stationary-states-para1}.
Since the passive part is composed of a single link, $R(\omega)$ is periodic of period $2\pi/T_2$. 
As $T\ll T_2$, the variation of $R(\omega)$ is slow compared over $\delta \omega = 2\pi/T$, so that
the distance between the active modes is approximately equal to $2\pi/ T$. 
Moreover, we see that the intensity drops in correspondence of the minima
of the transfer function, while the local maxima of the field intensity are 
located in the vicinity of the maxima of $\left|R(\omega)\right|$ as well as of the zeros
of the phase $\arg R(\omega)$; this corresponds to an optimal transfer
function for the minimal attenuation and no phase delay of the passive section. 

For larger $T_2$, the spacing is more irregular. In particular, if $T_2 \ll T$, it is the length
of the passive link, which determines the average mode spacing (see the middle row in
Fig.~\ref{fig:Stationary-states-para1}).

\section{Conclusions and future perspectives}

In this paper we have introduced a formalism for the study of
networks whereby waves with different frequencies propagate and interfere
with one another, according to resonances implicitly selected by the
network structure. As a result, we are able to study physical networks composed
of both active and passive links, where the waves can be either amplified
or damped. The dynamical system is represented as an abstract network whose nodes correspond to
the electric fields propagating along the physical links, while its
links correspond to the splitters which couple incoming with outgoing
fields.

This formalism can be considered as an extension of the method developed
to analyse quantum graphs to a context where wave propagation is nonlinear and in the presence of 
both gain and losses.
Under the simplifying assumption of a linear damped propagation along
the passive links, the corresponding dynamics can be treated as delayed boundary conditions. 

In this paper we restricted the analysis to setups composed of a single active link. In such cases, the action
of the passive subnetworks can be schematized through the action of a possibly complex transfer function,
which contributes to select the active degrees of freedom, in principle, ``offered" by the amplification
along the single active link. The potential high dimensionality of the resulting dynamics is ensured by the
delayed character of the equations.
In the presence of multiple active links, we expect additional peculiarities to emerge because of the
interactions between them. This challenging task goes, however, beyond the scope of the present work, which
is mostly methodological. 

From the point of view of the model, a relevant
assumption that has helped simplifying its mathematical structure
is the adiabatic elimination of the atomic polarization. 
This is a legitimate approximation in the
context of semiconductor lasers, but much less so for, e.g. erbium
doped optical fibers. 
A relevant step forward would be the extension
of the formalism to such systems. This objective can in principle
be achieved by explicitly including the spatial dependence
along the active links; however the corresponding computational complexity
would be so high as to make their analysis practically unfeasible.
The true question is indeed under which conditions it is possible
to keep a \textit{spaceless} mathematical structure, integrating out
the spatial dependence of the fields. This is a nontrivial step. Already
in the context of semiconductor lasers, it should be noted that our
equation of reference, used to describe an active link (see Eq.~(\ref{eq:active link}))
is partially phenomenological. As shown by Vladimirov and Turaev \citep{Vladimirov2005},
the spatial integration of the population inversion leads to an algebraic
delay equation, which exhibits time singularities under the form of
high frequency instabilities. The proposed solution, also adopted
herewith consists in the introduction of a filter (schematized by
the time derivative of the field and identified by the bandwidth $\beta$),
which regularizes the overall evolution.

This problem is reminiscent of the difficulty which emerges in a relatively
similar class of models proposed more than ten years ago to describe
an ensemble of dynamical units (the network nodes) which perform instantaneous
Boolean operations in the absence of an external clock which synchronizes
the single operations (see Ghil et al.~\citep{GhilZaliapinColuzzi2008}).
The resulting mathematical model consists of a set of standard (non-differential)
delay equations, where, like here, the delays originate from the transmission
times along the single links. In such a context, an unphysical ultraviolet
divergence appears as a consequence of the assumption of the instantaneous
response of the single devices. Also in that case, the evolution equation
has been phenomenologically regularized by turning it into a differential
equation.

Besides the extension of our formalism to a still wider class of optical
networks, another direction worth exploring is the actual dynamical
regimes that can arise within the current context. In this paper,
we have limited ourselves to determining the (many) stationary solutions.
What about their stability and the onset of irregular dynamics
because of the mutual nonlinear interactions? 
Finally, useful information and hints can instead arise from the analysis of
a simpler setup: the zero-delay limit, as the LANER reduces to a 
network of 3-d coupled oscillators with phase-shift symmetry: 
it would be interesting to see to what 
extent its dynamics differs from that of similar dynamical systems, more
extensively studied in the literature.

\section*{Acknowledgements}
Funding: SY was supported by the German Science Foundation (Deutsche Forschungsgemeinschaft, DFG) [project No.~411803875]

\end{document}